\documentclass[epj]{svjour}
\usepackage{graphics}
\usepackage{amsmath,amssymb,bbm}
\usepackage{subeqnarray}

\usepackage{color}

\definecolor{darkmagenta}{rgb}{.8,0,.8}
\definecolor{darkred}{rgb}{.7,0,0}
\definecolor{darkblue}{rgb}{0,0,.7}
\definecolor{darkgreen}{rgb}{0,.6,0}
\definecolor{normblue}{rgb}{0,0,1}

\newcommand{\nn}{\nonumber}
\newcommand{\U}{\Upsilon}

\newcommand{\be}{\begin{equation}}
\newcommand{\ee}{\end{equation}}
\newcommand{\bea}{\begin{eqnarray}}
\newcommand{\eea}{\end{eqnarray}}
\newcommand{\brr}{\begin{array}}
\newcommand{\err}{\end{array}}

\def\lan{\langle}

\def\rar{\to}

\def\ga{\gamma}
\def\te{\theta}

\def\om{\omega}

\def\mass{{_{1,2}}}
\def\1{{_{1}}}\def\2{{_{2}}}

\def\lsim{\hbox{ \raise.35ex\rlap{$<$}\lower.6ex\hbox{$\sim$}\ }}
\def\gsim{\hbox{ \raise.35ex\rlap{$>$}\lower.6ex\hbox{$\sim$}\ }}
\def\slashed{\partial\hspace{-6pt}\slash}

\begin{document}


\title{Doubling of the Algebra and Neutrino Mixing\\
within
Noncommutative Spectral Geometry}
\subtitle{Doubling of the Algebra and Neutrino Mixing}

\author{Maria Vittoria Gargiulo
\footnote{mariavittoria.gargiulo@gmail.com}
\inst{1}  \and Mairi
Sakellariadou
\inst{2}
\and
Giuseppe Vitiello
\inst{1}}
\institute
{Dipartimento di Fisica, I.N.F.N., Universit\`a di Salerno,
I-84100 Salerno, Italy
\and
Department of Physics, King's College London,
University of London, Strand WC2R 2LS, London, U.K.}

\date{ 2 December 2013 \and ...}

\abstract
{We study physical implications of the doubling of the algebra, an
essential element in the construction of the noncommutative spectral
geometry model, proposed by Connes and his collaborators as offering a
geometric explanation for the standard model of strong and electroweak
interactions.  Linking the algebra doubling to the deformed Hopf
algebra, we build Bogogliubov transformations and show the emergence
of neutrino mixing.}

\PACS{
     {02.40.Gh}{Noncommutative geometry}
\and
 {14.60.Pq}{Neutrinos mass and mixing}
    \and
{12.10.Dm}{Unified theories and models of strong and electroweak interactions}
}

\authorrunning{M.\ V.\ Gargiulo, M.\ Sakellariadou, G.\ Vitiello}
\titlerunning{Doubling of the Algebra and Neutrino Mixing within NCSG}

\maketitle
\section{Introduction}
Approaching the Planck energy scale one expects that the notion of a
continuous geometrical space ceases to be valid. At such high energy
scales the simple hypothesis that physics can be described by the sum
of the Einstein-Hilbert action and the Standard Model (SM) action can
no longer be valid. The Noncommutative Spectral Geometry (NCSG)
model~\cite{ncg-book1,ncg-book2} treats the SM as a low-energy
phenomenological model which however dictates the geometry of
spacetime at high energy scales.  Hence, the aim of NCSG is to reveal
the small-scale structure of spacetime from our knowledge of
experimental particle physics at the electroweak scale.  Following this
approach, implies that to construct a quantum theory of gravity
coupled to matter, we will consider that the gravity-matter interaction incorporates the most crucial aspect of the dynamics.

At very high energy scales Quantum Gravity could imply that spacetime
is a strongly noncommutative manifold.  For energies a few orders of
magnitude below the Planck scale, however, it is conceivable to
consider that the algebra of coordinates can be given by a slightly
noncommutative algebra~\cite{ncg-book1,ncg-book2,Dungen} which, if
appropriately chosen, can lead to the SM coupled to
gravity~\cite{ccm,ac2006,cchiggs}.  This slightly noncommutative manifold has been
chosen to be the tensor product of an {\sl internal}
(zero-dimensional) Kaluza-Klein (discrete) space and a continuous
(four-dimensional) spacetime. Thus, geometry close but below the
Planck scale is defined by the product ${\cal M}\times {\cal F}$ of a
continuum compact Riemannian manifold ${\cal M}$ (for the spacetime)
and a discrete finite noncommutative space ${\cal F}$ (for the SM)
composed by only two points; such a geometry is called {\sl almost
commutative}.

This choice of the doubling of the algebra,
which can be interpreted as considering a geometric space formed by
two copies (branes) of a four-dimensional manifold, has deep physical
implications. As pointed out in Ref.~\cite{Sakellariadou:2011wv}, the
doubling of the algebra is required in order to accommodate gauge
symmetries, which are necessary to describe the SM, while the doubling of the
algebra is also related to dissipation, hence to information loss,
thus containing the seeds of quantisation.

The purpose of this paper is to show that the doubling of the algebra
is also the main element to explain neutrino mixing. Hence, our main motivation is to provide a physical meaning to the mathematical construction of NCSG, a model constructed to give a purely geometric explanation of the SM. The issue of neutrino mixing is particularly important, since it
opens interesting perspectives on the physics beyond the
SM. Many experimental efforts are thus pursued~\cite{neutrinoExperim}, and
the quantum field theory for neutrino mixing (and, in general, for
particle mixing) has been
formulated~\cite{threeNeutrinos,neutrinoMixQFT,Palmer},  providing the framework for various studies, also in conjunction with dark energy and dark matter scenarios~\cite{neutrinoDark}.

In what follows,
we first give in Section~\ref{intro} a brief presentation of the NCSG
elements that we will then use. We then summarise in
Section~\ref{neutrinos} how neutrinos appear within this
construction. In Section~\ref{Hopf} we relate the algebra doubling,
which is a crucial element of the NCSG model, to the Hopf
noncommutative algebra and Bogogliubov transformations. In
Section~\ref{mix} we show how the doubling of the algebra implies
neutrino mixing. We then close with our conclusions in
Section~\ref{conclusions}.

\section{Elements of NCSG}
\label{intro}
Noncommutative spectral geometry,  as an approach to unification, is based on three {\sl ansatz}, which we state below:

$\bullet$ At some energy level, close but below the Planck scale,
geometry is described by the product of a four-dimensional smooth
compact Riemannian manifold, ${\cal M}$, with a fixed spin structure by
a discrete noncommutative space, ${\cal F}$, composed by only two
points.\\ The noncommutativity of ${\cal F}$ can be expressed by a real
spectral triple $({\cal A}_{\cal F}, {\cal H}_{\cal F},
D_{\cal F})$, where ${\cal A}_{\cal F}$ is an involution of operators
on the finite-dimensional Hilbert space ${\cal H_F}$ of Euclidean
fermions, and ${\cal D_F}$ is a self-adjoint unbounded operator in
${\cal H}_{\cal F}$. The algebra ${\cal A}_{\cal F}$ contains all
information usually carried by the metric. The axioms of the spectral
triples imply that the Dirac operator of the internal space, $D_{\cal
F}$, is the fermionic mass matrix. The Dirac operator is the inverse
of the Euclidean propagator of fermions. The spectral geometry
for ${\cal M}\times {\cal F}$ is thus given by
\bea &\mathcal{A}=C^\infty (\mathcal{M}) \otimes \mathcal{A_F} =
C^\infty ( \mathcal{M , A_F})~,\nonumber\\ &\mathcal{H} =
\mathcal{L}^2(\mathcal{ M }, S) \otimes \mathcal{H_F} =
\mathcal{L}^2(\mathcal{M} , S \otimes \mathcal{H_F ) }~\nonumber\\ & D =
D_{\mathcal{M}} \otimes 1 + \gamma_5 \otimes D_{
\mathcal{F}}~,\nn \eea
where $C^\infty (\mathcal{M,C})$ is the algebra of smooth complex
valued functions on ${\cal M}$; $\mathcal{L}^2(\mathcal{ M }, S)$ is
the space of square integrable Dirac spinors over ${\cal M}$; $D_{\cal
M}$ is the Dirac operator $\slashed_{\cal
M}=\sqrt{-1}\gamma^\mu\nabla_\mu^s$ on ${\cal M}$; and $\gamma_5$ is
the chirality operator in the four-dimensional case.

$\bullet$ The finite dimensional algebra ${\cal A_F}$, which is the main
input, is chosen to be~\cite{Chamseddine:2007ia}
\begin{equation}
\mathcal{A_F}=M_{a}(\mathbb{H})\oplus M_{k}(\mathbb{C})~,
\end{equation}
with $k=2a$ and $\mathbb{H}$ being the algebra of
quaternions.  This choice was made due to the three
following reasons: (i) the model should account for massive neutrinos
and neutrino oscillations so it cannot be a left-right symmetric
model, like for instance $\mathbb{C}\oplus\mathbb{H}_{\rm
L}\oplus\mathbb{H}_{\rm R}\oplus M_3(\mathbb{H})$; (ii) noncommutative
geometry imposes constraints on algebras of operators in the Hilbert
space; and (iii) one should avoid fermion doubling. The first possible
value for the even number $k$ is 2, corresponding to a Hilbert space
of four fermions, but this choice is ruled out from the existence of
quarks. The next possible value is $k=4$ leading to the correct number
of $k^2=16$ fermions in each of the three generations. This is the
most economical choice~\cite{Devastato:2013oqa} that can account for the SM.

$\bullet$ The action functional is dictated by the spectral action
principle, which affirms that the bosonic part of the action
functional depends only on the spectrum of the Dirac operator ${\cal
D}$ and is of the form
\begin{equation}
{\rm Tr}\left(f\left(\frac {{\cal D}}{\Lambda} \right)\right)~,
\label{trace}
\end{equation}
where $f$ is a positive even function of the real variable and it
falls to zero for large values of its argument, while the parameter
$\Lambda$ fixes the energy scale. Thus, the action functional sums up
eigenvalues of the Dirac operator which are smaller than the cut-off
scale $\Lambda$.  Since the bosonic action only depends on the
spectrum of the line element, i.e. the inverse of the Dirac operator,
the operator ${\cal D}$ contains all information about the
bosonic part of the action.

The trace, Eq.~(\ref{trace}), is then evaluated with heat kernel
techniques and is given in terms of geometrical Seeley-deWitt
coefficients $a_n$. Since $f$ is a cut-off function, its Taylor
expansion at zero vanishes. Therefore, its asymptotic expansion
depends only on the three momenta $f_0$, $f_2$ and $f_4$, which are
related to the coupling constant at unification, the gravitational
constant and the cosmological constant, respectively.  In this sense,
the choice of the test function $f$ plays only a limited r\^ole.
Hence,
\begin{equation}
{\rm Tr}\left(f\left(\frac {\cal D}{\Lambda} \right)\right ) \sim 2
\Lambda^4 f_4 a_0 + 2 \Lambda^2 f_2 a_2 + f_0 a_4~,
\end{equation}
where
\be
f_k=\int_0^\infty f(u)u^{k-1} du~.
\nonumber
\ee
The gravitational Einstein action is thus obtained by the expansion of
the action functional.\\  The coupling with fermions is obtained by
adding to the trace, Eq.~(\ref{trace}), the term
\be
{\rm Tr}\frac{1}{2}\langle J\psi,{\cal D}\psi\rangle
~,
\ee
where $J$ is the real structure on the spectral triple and $\psi$ is
an element in the space ${\cal H_F}$.

In the presence of gauge fields $A$, there is a modification in the
metric (within noncommutative geometry, one does not focus on
$g_{\mu\nu}$ but on the Dirac operator instead), leading to the inner
fluctuations of the metric 
\be \label{5}
{\cal D}\rightarrow{\cal D}_A={\cal D}+A+\epsilon'JAJ^{-1}~,
\ee
where $A$ is a self-adjoint operator of the form
\be
A=\sum_{j}a_j[D,b_j]~,~~ a_j, b_j\in {\cal A}~,\nn
\ee
$J$ is an antilinear isometry and $\epsilon'\in \{-1,1\}$.  Applying
the action principle to ${\cal D}_A$ one obtains the combined
Eistein-Yang-Mills action.  Thus, the fermions of the SM provide the
Hilbert space of a spectral triple for a suitable algebra, while the
bosons arise as inner fluctuations of the corresponding Dirac
operator.

In conclusion, the full Lagrangian of the SM minimally coupled to
gravity, is obtained as the asymptotic expansion (in inverse powers of
$\Lambda$) of the spectral action for the product geometry ${\cal
M}\times {\cal F}$. This geometric model can explain the SM
phenomenology~\cite{ccm,cchiggs,Devastato:2013oqa,Chamseddine:2013rta}. Moreover, since this
model lives by construction at very high energies, it can provide a
natural framework to address early universe cosmological
issues~\cite{Nelson:2008uy,Nelson:2009wr,Marcolli:2009in,mmm,Nelson:2010ru,Nelson:2010rt,Sakellariadou:2010nr,Sakellariadou:2011dk,Lambiase:2013dai}.

\section{Neutrinos within the NCSG model}\label{neutrinos}
In the context on NCSG, neutrinos appear naturally as Majorana spinors
(so that neutrinos are their own antiparticles), for which
the mass terms in the Lagrangian can be written as
\be \frac{1}{2}\sum_{\lambda\kappa}\bar\psi_{\lambda L}{\cal
S}_{\lambda\kappa}\hat\psi_{\kappa R}
+\frac{1}{2}\sum_{\lambda\kappa} \overline{\bar\psi_{\lambda L}{\cal
S}_{\lambda\kappa}\hat\psi_{\kappa R}}~,\nn \ee
where the subscript $_{L,R}$ stand for left-handed, right-handed
states, respectively. The off-diagonal parts of the symmetric matrix
${\cal S}_{\lambda\kappa}$ are the Dirac mass terms, while the
diagonal ones are the Majorana mass terms.

Within NCSG, one can show~\cite{ncg-book2} the existence of a Dirac
operator ${\cal D_F}$ for the algebra
\bea {\cal A_F}&=&\{(\lambda, q_L,\lambda, m)|\lambda\in \mathbb{C},
q_L\in\mathbb{H}, m\in M_3(\mathbb{C}) \} \nn\\
&\sim& \mathbb{C}\oplus \mathbb{H}\oplus M_3(\mathbb{C})~,\nn\eea
with off-diagonal terms. In particular, one can show~\cite{ncg-book2}
that there exist $3\times 3$ matrices (3 for the number of
generations) $\U_e, \U_\nu, \U_d, \U_u $ and a symmetric $3\times 3$
matrix (3 for the number of generations) $\U_R$, such that ${\cal D_F}$ is of
the form
\be
{\cal D_F}(\Upsilon)=\left( \begin{array}{ccc}
S & T^\star \\
T & \overline{S} \end{array}\right)~.
\ee
$S$ is a linear map
\be
S=S_l\oplus(S_q\otimes 1_3)~,\nn
\ee
with $1_3$ the identity $3\times 3$ matrix and
\be
S_l=\left( \begin{array}{cccc}
0 & 0 & \Upsilon_\nu^\star & 0 \\
0 & 0 & 0 & \Upsilon_e^\star  \\
\Upsilon_\nu & 0 & 0 & 0 \\
0 & \Upsilon_e & 0 & 0  \end{array}\right)~,
~
S_q=\left( \begin{array}{cccc}
0 & 0 & \Upsilon_u^\star & 0 \\
0 & 0 & 0 & \Upsilon_d^\star  \\
\Upsilon_u & 0 & 0 & 0 \\
0 & \Upsilon_d & 0 & 0  \end{array}\right)~;\nn
\ee
the subsripts $_q$ and $_l$ stand for quarks and leptons,
respectively.  The $^\star$ denotes adjoints and $\bar S=\bar
S_l\oplus(1_3\otimes \bar S_q)$ act on ${\cal H}_{\bar f}$
by the complex conjugate matrices; we have splitted ${\cal H_F}$
according to ${\cal H_F}={\cal H}_f\oplus {\cal H}_{\bar f}$. Finally, $T$ a
is linear map, so that $T(\nu_R)=\U_R\bar\nu_R$.

The presence of the symmetric matrix $\U_R$ in the Dirac operator of
the finite geometry ${\cal F}$ accounts for the Majorana mass terms,
while $\U_\nu$ is the neutrino Dirac mass matrix. Hence, the
restriction of ${\cal D_F}(\U)$ to the subspace of ${\cal H_F}$ with the  $(\nu_R,
\nu_L,\bar\nu_R,\bar\nu_L)$  basis can be written as a matrix~\cite{ncg-book2}
\be
\left( \begin{array}{cccc}
0 & M_\nu^\star & M_R^\star & 0 \\
M_\nu & 0 & 0 & 0 \\
M_R & 0 & 0 & \bar M_\nu^\star  \\
0 & 0 & \bar M_\nu & 0
  \end{array}\right)~,
\ee
where $M_\nu=(2M/g)K_\nu$ with
\be 2M=\Big[\frac{{\rm Tr}(\U_\nu^\star \U_\nu+\U_e^\star \U_e
+3(\U_u^\star \U_u+\U_d^\star \U_d)}{2} \Big]^{1/2}~, \ee
$K_\nu$ the neutrino Dirac mass matrix and $M_R$ the Majorana mass matrix.

The equations of motion of the spectral action imply that the largest
eigenvalue of $M_R$ is of the order of the unification scale. The
Dirac mass $M_\nu$ turns out to be of the order of the Fermi energy, thus much
smaller. In conclusion, the way the NCSG model has been built, it can account
for neutrino mixing and the seesaw mechanism.

In the next section we will discuss the links between the NCSG
doubling of the algebra and the deformed Hopf algebra and we will show
how to obtain the Bogoliubov transformations from linear combinations
of deformed coproducts in the Hopf algebra. The neutrino mixing in the
context of NCSG will be then discussed in Section V. Mixing will
appear to be implied by the doubling of the algebra which is the core
of Connes construction. The neutrino mixing thus appears to be a
manifestation of the spectral geometry nature of the construction.

\section{Algebra doubling, Hopf noncommutative algebra and  Bogoliubov
transformations}\label{Hopf}
Let us consider~\cite{ncg-book2} the finite geometry ${\cal F}$ described by
\be
{\cal F}=({\cal A_F}, {\cal H_F}, D_{\cal F}, J_{\cal F}, \gamma_{\cal F})~,\nonumber
\ee
where $J_{\cal F}$ is an antilinear isometry and $\gamma_{\cal F}$ is the 
$\mathbb{Z}/2$-grading on ${\cal M_F}$. The pair $(J_{\cal F}, \gamma_{\cal F})$ satisfies
\be
J_{\cal F}^2=1 ~,~ J_{\cal F}\gamma_{\cal F}=-\gamma_{\cal F}J_{\cal F}~.\nonumber
\ee
Then consider the product of the finite noncommutative geometry  ${\cal F}$, with the spectral triple associated to the commutative geometry of a compact four-dimensional Riemannian spin manifold of spacetime ${\cal M}$. 
Note that for a compact spin four-manifold ${\cal M}$, the associated spectral triple is $(C^\infty ({\cal M}), L^2({\cal M},S),\slashed_{\cal M})$.
The product geometry ${\cal M}\times {\cal F}$ is the real spectral triple~\cite{ncg-book2} (see also Section~\ref{intro}) 
\bea
({\cal A}, {\cal H}, {\cal D},  J, \gamma)=\nonumber\\
(C^\infty ({\cal M}), L^2({\cal M},S),\slashed_{\cal M},J_{\cal M}, \gamma_5)\otimes
({\cal A_F}, {\cal H_F}, D_{\cal F}, J_{\cal F}, \gamma_{\cal F})\nonumber
\eea
defined as
\be\label{AcrosA}
({\cal A}, {\cal H}, {\cal D}, J, \ga) =({\cal A}_1, {\cal H}_1, {\cal
D}_{1}, J_1, \ga_1) \otimes ({\cal A}_2, {\cal H}_2, {\cal D}_2, J_2, \ga_2 )
\ee
with
\bea\label{double} {\cal A} &=& {\cal A}_1 \otimes {\cal A}_2 ~,~{\cal
H} = {\cal H}_1 \otimes {\cal H}_2~, \nonumber\\ {\cal D} &=& {\cal
D}_1 \otimes 1 + {\gamma}_1 \otimes {\cal D}_2~,\nonumber\\ \gamma &=&
{\gamma}_1 \otimes {\gamma}_2~,~~~ J= J_1 \otimes J_2~, \eea
and
\be \label{J} J^{2} = -1, ~~ [J,{\cal D}] = 0 , ~~[J_1,\gamma_1] = 0 ,~~\{J,\gamma\}= 0,
\ee
where square and curl brackets denote commutators and anticommutators, respectively.
Note that the resulting geometry $({\cal A, H}, D, J, \gamma)$ is of $KO$-dimension 10=2 modulo 8.
The difference between the two algebras ${\cal A}_1$ and ${\cal A}_2$ is that in one the multiplication is made ``row by column" while in the other one, multiplication is made ``column by row". These two algebras are related through the conjugation operator~\cite{ncg-book2}. 

The doubling map, given in
Eq.~(\ref{AcrosA}), which is the main element of Connes' NCSG, is intimately related to the noncommutative Hopf algebra characterised by the deformed coproduct map. This can be seen by introducing the (standard) notation 
\be
a \otimes 1 \equiv a_1~,~ 1 \otimes a \equiv a_2~,\nonumber
\ee
with
\be
\{a_i, a_j\} = 0 = \{a_i,a^{\dag}_j\}, ~ i\neq j, ~i,j = 1,2~,\nonumber
\ee
 and observing that the prescription to work in the NCSG two-mode
space ${\cal H} = {\cal H}_1 \otimes {\cal H}_2 $ is provided by the Hopf noncommutative coproduct operators given by~\cite{Celeghini:1998sy}
\bea \label{aq} \Delta a_q &=& a_q \otimes q^H +
q^{-H} \otimes a_q~,\nonumber\\
\Delta a^{ \dagger}_q &=& a^{ \dagger}_q \otimes q^H
+ q^{-H} \otimes a^{ \dagger}_q ~, \nonumber\\
\Delta H  &=&  H  \otimes \mathbbm{1}  +  \mathbbm{1} \otimes H ~,\nonumber\\
\Delta N  &=& N  \otimes  \mathbbm{1}  +  \mathbbm{1} \otimes N~,
\label{eqn:copr}
\eea
The noncommutative Hopf algebra is thus embedded in Connes' construction.
Note that noncommutativity is guaranteed by the so called ``deformation'' parameter $q$. The $H$ and $N$ are operators of the algebra (see below). In Eq.~(\ref{aq}), we have also used the notation of the $q$-deformed $h_q(1\mid 1)$ fermionic Hopf algebra for the operators $a_q$ and $a^{ \dagger}_q$. We indeed recall that the algebra $h(1\mid 1)$ is generated by the set of
operators $\{a , a^{ \dagger},H, N \} $ with
\bea
\{a , a^{\dagger} \} = 2H~,~~ [N , a] = -a~, ~~~ [N , a^{\dagger}] = a^{\dagger}~,
\eea
and $[H , \bullet] = 0$, with $H$ a central operator, constant in each
representation. The deformed algebra Hopf algebra $h_q(1\mid 1)$ embedded in Connes' construction is defined by
\be \label{qalg}
\{ a_q , a^{\dagger}_q \}= [2H]_q~,~~~
[ N , a_q ] = -a_q~,~~~
[ N , a^{\dagger}_q ] = a^{\dagger}_q~,
\ee
where $[H , \bullet] = 0$,
with $N_q\equiv N$ and $H_q\equiv H$,  while  $[x]_q$ is defined by
\be
[x]_q=\frac {q^x -q^{-x}} {q-q^{-1}}~ .
\ee
The Casimir operator $\mathcal{C}_q$ is given by $\mathcal{C}_q =
N[2H]_q - a^{\dagger}_q a_q$~. In the fundamental representation we
have $H = 1/2$ and the Casimir operator is thus zero, $\mathcal{C}_q =
0$. Note that the $q$-deformed coproduct definition is such that
$[{\Delta a_{q}, \Delta a^{\dag}_{q}}] = [2 \Delta H]_{q}$, etc.,
namely the $q$-coproduct algebra is isomorphic with the one defined by
Eq.~(\ref{qalg}).  Requiring $a, a^{\dagger}$ and $a_q, a^{\dagger}_q$
to be adjoint operators implies that $q$ can only be of modulus one,
hence $q\sim e^{i \Theta} $. In the fundamental
representation $h(1\mid 1)$ and $h_q(1\mid 1)$ coincide, as it happens
in the spin $1/2$ representation; the differences appearing only at
the level of the corresponding coproducts (and in the higher spin
representations). Also note that for consistency with the coproduct
isomorphism, the
Hermitian conjugation of the coproduct must be supplemented by the
inversion of the two spaces ${\cal H}_1 $ and ${\cal H}_2 $ in the
two-mode space ${\cal H}$.

In conclusion, we have  seen that the noncommutative ($q$-deformed) Hopf algebra is embedded in the NCSG construction whose central ingredient is the doubling map $\mathcal{A} \to \mathcal{A}_1 \otimes
\mathcal{A}_2 $.

We are now ready to show
that the $q$-deformed coproduct turns out to be related to the Bogoliubov
transformations, a key ingredient in the neutrino mixing formalism (see Section \ref{mix}) .  By resorting to the result of Ref.~\cite{Celeghini:1998sy}, let us define the operators $A_q$ and $B_q$, as 
\bea A_q
&\equiv&  \frac{\Delta a_q}{\sqrt{[2]_q}}
=\frac 1 {\sqrt{[2]_q}} (e^{i\Theta}a_1+e^{-i\Theta}a_2)~,\nn \\
B_q
&\equiv& \frac 1 {i \sqrt{[2]_q}} \frac \delta {\delta\Theta} \Delta a_q
=\frac 1 {\sqrt{[2]_q}} (e^{i\Theta}a_1-e^{-i\Theta}a_2)~, \eea
obtained from Eq.~(\ref{aq}) with $q=q(\Theta)\equiv e^{i2\Theta}$.
The anticommutation relations read
\bea \{A_q,A^{\dagger}_q\}=1~&,&~\{B_q,B^{\dagger}_q\}=1~,\nn\\
 \{A_q,B_q\}=0~&,&~\{A_q,B^{\dagger}_q\}= \tan{2\Theta}~.  \eea
Let us then construct the operators
\bea a(\Theta)&=&\frac{1}{\sqrt 2} \big( A(\Theta) +B(\Theta)
\big)~,\nn\\ \tilde{a}(\Theta)&=&\frac{1}{\sqrt 2} \big( A(\Theta)
-B(\Theta) \big)~,
\label{bogol}
\eea
where
\bea & A(\Theta)\equiv \frac{\sqrt{[2]_q}}{2\sqrt 2} \big[
A_{q(\Theta)} + A_{q(-\Theta)} + A^{\dagger}_{q(\Theta)} -
A^{\dagger}_{q(-\Theta)}\big]~, \nn\\ & B(\Theta)\equiv
\frac{\sqrt{[2]_q}}{2\sqrt 2} \big[ B_{q(\Theta)} + B_{q(-\Theta)} -
B^{\dagger}_{q(\Theta)} + B^{\dagger}_{q(-\Theta)}\big]
\eea
 Hence,
\bea a(\Theta)&=& U (\Theta) \, a_1  - i \, V (\Theta)\,a_2^{\dagger} ~,\nn\\
\tilde{a}(\Theta)&=& U (\Theta) \, a_2  + i\,V (\Theta)\,
a_1^{\dagger}~,
\label{bogola}
\eea
with 
\be
\{a(\Theta),\tilde{a}(\Theta)\}=0~,\nonumber
\ee
and
\be
U^2 (\Theta) + V^2 (\Theta) = 1~,
~ U (\Theta)= \cos \Theta~, 
~V (\Theta)= \sin \Theta~.
\nonumber
\ee
 The only  nonzero anticommutation relations are 
\be\{a(\Theta),a^{\dagger}(\Theta)\}=1~,~\{\tilde{a}(\Theta),
\tilde{a}^{\dagger}(\Theta)\}=1~.
\label{anticomm}
\ee 
Equation~(\ref{bogola}) is the Bogoliubov transformation of the
pair of creation and annihilation operators $(a_1,a_2)$ into 
$(a(\Theta),\tilde{a}(\Theta))$.
Equations~(\ref{bogol})-(\ref{bogola}) show that the
Bogoliubov-transformed operators, $a(\Theta)$ and $\tilde{a}(\Theta)$,
are linear combinations of the coproduct operators defined in terms of
the deformation parameter   $q(\Theta)$ and their
$\Theta$-derivatives. Notice in Eq.~(\ref{bogola}) the antilinearity
of the {\it tilde conjugation} $c {\cal O} \rightarrow c^{*} {\tilde
{\cal O}} $ which reminds of the antilinearity of the $J$ isometry
introduced in Section 2.~\footnote{For more details on this and other
features of the $q$-deformed Hopf algebra and the Bogoliubov
transformation, we refer the reader to Refs.~\cite{Celeghini:1998sy}
and \cite{BJV:2011}.}

It is worth noting that besides our discussion on neutrino mixing,
Bogoliubov transformations are also relevant for quantum aspects of
the theory. Indeed, they are known to describe the transition among
unitarily inequivalent representations of the canonical
(anti)commutation relations in quantum field theory (QFT) at finite
temperature and are therefore a key tool in the description of the
non-equilibrium dynamics of symmetry breaking phase
transitions~\cite{BJV:2011,Umezawa1983,LesHouchesVit}. Here
we have shown that Bogoliubov transformations are encoded in the very
same structure of the algebra doubling of Connes construction. This
links the NCSG construction with the nonequilibrium dynamics of the
early universe, as well as with elementary particle physics.

In the next section we show that the noncommutative Hopf algebra
embedded in the NCSG construction rules the neutrino mixing phenomenon
which is thus ``implied'' by the same construction.

\section {Neutrino mixing}
\label{mix}
Our aim here is to show how Bogoliubov transformations, and thus the noncommutative Hopf algebraic structure which has been shown above to be embedded in the NCSG construction, may explain neutrino mixing. Hence, neutrino mixing can find its natural setting in the NGSG construction.
Our discussion is based on the quantum field theory algebraic structure and is therefore of general character, regarless of the number of Euclidean dimensions. Thus, our result applies to Dirac neutrinos~\cite{neutrinoMixQFT} as well as to Majorana neutrinos~\cite{Palmer}, and in principle to other particle mixing (such as meson mixing and quark mixing), too~\cite{Connes:2008wi}.  For concreteness, we refer below to Majorana neutrinos~\cite{Palmer}.

In the context of NCSG, neutrinos appear naturally as Majorana spinors (neutrinos are their own antiparticles). Connes and his collaborators have derived their model after a Wick rotation to Euclidean signature~\cite{ncg-book2}. Since, unlike Dirac spinors, Majorana spinors do not have a Euclidean version, one may think that there is a problem in the context of NCSG model  for massive neutrinos. However, as discussed in detail in Ref.~\cite{ncg-book2},  one can use a formalism based on the Pfaffian and Grassmann variables to obtain a
substitute for the formalism of Majorana spinors in the Euclidean setup.

Let us introduce the Lagrangian
\bea L(x)&=&\bar{\psi}_m(x)(i \slashed - M_{\rm d})\psi_m (x)
\nonumber\\ &=&\bar{\psi}_f(x)(i \slashed -M)\psi_f (x)~, \eea where
we use the notation $x \equiv ({\bf x},t)$, while
$\psi^T_m=(\nu_1,\nu_2)$ denote the neutrino fields with nonvanishing
masses $m_1$ and $m_2$, respectively, and $\psi^T_f=(\nu_e,\nu_\mu)$
stand for the flavor neutrino fields. We denote $M_{\rm d}=~{\rm
diag}(m_1 , m_2)$ and
$$M=\begin{pmatrix} m_e & m_{e\mu} \\ m_{e\mu} & m_{\mu} \end{pmatrix}~, $$
the mass matrices.  For simplicity, we consider only two neutrinos;
extension to three neutrino fields can be easily
done~\cite{threeNeutrinos}.  The mixing transformations connecting the
flavor fields $\psi_f$ to the fields $\psi_m$ are
\bea
& \nu_e (x)= \nu_1(x) \cos{\theta} + \nu_2 (x) \sin{\theta}~,\nn\\
& \nu_{\mu} (x)= - \nu_1(x) \sin{\theta} + \nu_2 (x) \cos{\theta}~.
\label{mixA}
\eea
The field quantization setting is the standard one; the $\psi_m$
fields are free fields in the Lehmann- Symanzik- Zimmermann (LSZ)
formalism of QFT and their explicit expressions in terms of creation
and annihilation operators $\alpha$ and $\alpha^\dag$ are
\begin{equation}
\nu_i (x) =\sum_{r=1,2} \int \frac {d^3 \mathbf{k}}{(2\pi)^\frac 3 2 }
e^{i \mathbf{k\cdot
x}}\big[u^r_{\mathbf{k},i}(t)\alpha^r_{\mathbf{k},i}+
v^r_{-\mathbf{k},i}(t)\alpha^{r\dagger}_{-\mathbf{k},i}\big]~,
\end{equation}
where $u^r_{\mathbf{k},i}(t)=e^{-i \omega_{\mathbf{k},i}t}
u^r_{\mathbf{k},i}$ , $v^r_{\mathbf{k},i}(t)=e^{i
\omega_{\mathbf{k},i}t} v^r_{\mathbf{k},i}$, while $r$ is the helicity
index and $\omega_{\mathbf{k},i}=\sqrt{\mathbf{k}^2+m^2_i}$ with
$i=1,2$.  Note that the operator anticommutation relations and the
spinor wavefunctions orthogonality and completeness relations are the
standard ones and we do not report them here for brevity.

Let $G_{\theta}(t)$ denote the generator of the field mixing transformations
Eq.~(\ref{mixA}):
\bea
& \nu_e (x)= G^{-1}_{\theta}(t)\nu_1(x) G_{\theta}(t)~,\nn\\
& \nu_{\mu} (x)= G^{-1}_{\theta}(t)\nu_2(x) G_{\theta}(t)~.
\eea
It is given by
\begin{equation}
G_{\theta}(t)=\exp{\Big[\frac \theta 2 \int d^3\mathbf{x} \big(
\nu^{\dagger}_1(x) \nu_2(x) - \nu^{\dagger}_2(x) \nu_1(x)
\big)}\Big]~.
\end{equation}
Due to the canonical anticommutation rules one can write
$G_{\theta}(t)=\textstyle\prod_{\mathbf{k}}
G^{\mathbf{k}}_{\theta}(t)$. Moreover, in the reference frame where
$\mathbf{k} =( 0 , 0 , |\mathbf{k}| )$, we have
$G^{\mathbf{k}}_{\theta}(t)=\textstyle\prod_r G^{\mathbf{k},
r}_{\theta}(t)$, with
\bea \label{GUV} G^{\mathbf{k}, r}_{\theta}(t)=\rm{exp}\Big\{ \theta
\Big[ U^\ast_{\mathbf{k}}(t) \alpha^{r\dagger}_{\mathbf{k},1}
\alpha^{r}_{\mathbf{k},2} - U_{\mathbf{k}}(t)
\alpha^{r\dagger}_{-\mathbf{k},2} \alpha^{r}_{-\mathbf{k},1}\nn\\
-\epsilon^r V^\ast_{\mathbf{k}}(t) \alpha^{r}_{-\mathbf{k},1}
\alpha^{r}_{\mathbf{k},2} +\epsilon^r V_{\mathbf{k}}(t)
\alpha^{r\dagger}_{\mathbf{k},1} \alpha^{r\dagger}_{-\mathbf{k},2}
\Big]\Big\}, \eea
where $\epsilon^r = (-1)^r$  and
\bea
U_{{\bf k}}(t)&\equiv& |U_{{\bf k}}|\,e^{i(\om_{k,2}-\om_{k,1})t}~,\nonumber\\
V_{{\bf k}}(t)&\equiv&|V_{{\bf
k}}|\,e^{i(\om_{k,2}+\om_{k,1})t}~.
\eea
For our purpose it is not essential to give here the explicit
expression of $|U_{{\bf k}}|$ and $|V_{{\bf k}}|$; the important point
is that
\be |U_{{\bf k}}|^{2}+|V_{{\bf k}}|^{2}=1~, 
\label{sincos}\ee 
which guarantees
that the mixing transformations preserve the canonical anticommutation
relations, i.e. they are canonical transformations.  Equation (\ref{sincos}) shows that one can always put $|U_{{\bf k}}|^{2} \equiv \cos^{2} \Theta_{{\bf k}}$ and $|V_{{\bf k}}|^{2} \equiv \sin^{2} \Theta_{{\bf k}}$.
Using Eq.~(\ref{GUV}) we define the flavor annihilation operators:
\bea 
\label{opcrea} 
 \alpha^r_{\mathbf{k},e}&\equiv&
G^{-1}_{\theta}\alpha^r_{-\mathbf{k},1} G_{\theta}(t)\nn\\
&=&\cos{\theta}\alpha^r_{\mathbf{k},1} + \sin{\theta}
\big(U^\ast_{\mathbf{k}}(t)\alpha^r_{\mathbf{k},2} +\epsilon^r
V_{\mathbf{k}}(t)\alpha^{r\dagger}_{-\mathbf{k},2} \big),\nn\\
\alpha^r_{\mathbf{k},\mu}&\equiv&
G^{-1}_{\theta}\alpha^r_{-\mathbf{k},2} G_{\theta}(t)\nn\\
&=&\cos{\theta}\alpha^r_{\mathbf{k},2} - \sin{\theta}
\big(U^\ast_{\mathbf{k}}(t)\alpha^r_{\mathbf{k},1} +\epsilon^r
V_{\mathbf{k}}(t)\alpha^{r\dagger}_{-\mathbf{k},1} \big) \eea
and similar relations for the flavor creation operators.

Note that the Bogoliubov coefficients $U_{{\bf k}}$ and $V_{{\bf k}}$ are  related to the noncommutative coproduct maps discussed in Section \ref{Hopf} (cf., e.g., Eq.~(\ref{bogola})). In this connection, we remark that the noncommutative coproduct maps are  related, {\sl not} to the mixing angle $\theta$, but to the Bogoliubov angles $\Theta_{{\bf k}}$. Moreover, inspection of Eq.~(\ref{opcrea}) shows that the mixing transformations
for the creation and annihilation operators produce ``nested''
operator rotation and time-dependent Bogoliubov transformations with
coefficients $U_{\mathbf{k}}(t)$ and $V_{\mathbf{k}}(t)$.  Since
deformed coproducts are a basis of Bogoliubov transformations, we have
thus shown that the field mixing ultimately rests on the algebraic
structure of the deformed coproduct in the noncommutative Hopf algebra
embedded in the algebra doubling of NCSG. Indeed, for vanishing value of $|V_{{\bf k}}|$, i.e. for vanishing $\Theta_{{\bf k}}$ for any ${\bf k}$, and thus $|U_{{\bf k}}|^{2} = 1$, there is only the field rotation (cf. Eqs.~(\ref{GUV}) and (\ref{opcrea})), {\it not} the mixing phenomenon.
Of course, the field rotation in the plane $\nu_1 - \nu_1$ is a unitary transformation out of which no ``new'' quantum number, such as the flavor (lepton) number of $\nu_e$ and $\nu_\mu$, can be generated, as instead it happens in the field mixing case. This result, as already said above, holds for the mixing of any considered particle, Dirac and Majorana neutrinos, quark or meson mixing.

We can finally express the flavor fields in terms of these flavor annihilation
and creation operators as~\cite{threeNeutrinos,neutrinoMixQFT}
\begin{equation}
\nu_\sigma (x) =\sum_{r=1,2} \int \frac {d^3 \mathbf{k}}{(2\pi)^\frac
3 2 } e^{i \mathbf{k\cdot
x}}\big[u^r_{\mathbf{k},j}(t)\alpha^r_{\mathbf{k},\sigma}+v^r_{-\mathbf{k},j}(t)
\alpha^{r\dagger}_{-\mathbf{k},\sigma}\big]\quad,
\end{equation}
with $\sigma , j = (e, 1),(\mu , 2)$ .

The flavor vacuum annihilated by the $\alpha^r_{\mathbf{k},\sigma}$,
$\sigma = e, \mu$, operators is defined by the action of the mixing
generator on the vacuum $|0\rangle_{1,2}$ annihilated by the
$\alpha^r_{\mathbf{k},i}$, $i=1,2$, operators
($\alpha^r_{\mathbf{k},1} |0\rangle_{1,2} = 0 =
\alpha^r_{\mathbf{k},2} |0\rangle_{1,2} $) as
\begin{equation}
|0(\theta,t)\rangle_{e,\mu}\equiv G^{-1}_{\theta}(t)|0\rangle_{1,2}.
\end{equation}
The expectation value of the number operator
$\alpha^{r\dagger}_{\mathbf{k},i}\alpha^r_{\mathbf{k},i}$, $i = 1,2$,
in such a vacuum state $|0(\theta,t)\rangle_{e,\mu}$ is nonzero, i.e.
\begin{equation}
_{e,\mu}\langle{0 (t)} |\alpha^{r\dagger}_{\mathbf{k},i} \alpha^{r}_{\mathbf{k},i}
| {0 (t)} \rangle_{e,\mu} =
|V_{\mathbf{k}}(t)|^{2} \,\sin^{2}(\theta) , ~~i = 1,2,
\end{equation}
which expresses that the flavored vacuum is a condensate (of couples)
of $i$-neutrinos, $i = 1,2$, hence its nonperturbative nature. 
We see that the expectation value of the number operator
vanishes in the $|V_{\mathbf{k}}(t)| \rar 0$ limit, i.e. in the
commutative limit where the Bogoliubov transformations are eliminated
(cf. Eqs.~(\ref{opcrea})).  We remark that the space of the neutrino
flavored states is unitarily inequivalent to the space of the mass
neutrino eigenstates.  Indeed, in the limit of the volume $V$ going to
infinity, one obtains
\bea\label{ineq} _\mass\lan 0|0 (t)\rangle_{e,\mu} \rar \, 0 ~, ~ {\rm
as}~ V \rar \infty \, ~~~{\rm for ~ any} ~t, \eea
which shows that $|0(t)\rangle_{e,\mu}$ and $|0(t)\rangle_\mass$ are
unitarily inequivalent representations of the canonical anticommutator
relations. In the absence of mixing ($\te=0$ and/or $m_1=m_2$) the
orthogonality between $|0(t)\rangle_{e,\mu}$ and $|0(t)\rangle_\mass$
disappears. Equation~(\ref{ineq}) can only hold in the QFT framework;
since there unitarily inequivalent representations exist, contrarily
to what happens in Quantum Mechanics (QM) where the von Neumann
theorem states the unitary equivalence of the representations of the
canonical anticommutation relations. Equation~(\ref{ineq}) also
expresses the nonperturbative nature of the field mixing mechanism.

The single (mixed) particle flavored state is given by
\begin{equation}
|\alpha^r_{\mathbf{k},\sigma}(t)\rangle \equiv
 \alpha^{r\dagger}_{\mathbf{k},\sigma} (t)|0(t)\rangle_{e,\mu}=
 G^{-1}_{\theta}(t) \alpha^{r\dagger}_{\mathbf{k},i}|0\rangle_{1,2}~,
\end{equation}
where $\sigma, i = e,1$ or $\mu,2$ .  States with particle number
higher than one are obtained similarly by operating repeatedly with
the creation operator $\alpha^{r\dagger}_{\mathbf{k},\sigma}$.  The
momentum operator for the free fields is
\begin{equation}
\mathbf{P}_i = \sum_{r=1,2} \int d^3 \mathbf{k} \, \mathbf{k} \Big(
\alpha^{r\dagger}_{\mathbf{k},i} \alpha^{r}_{\mathbf{k},i} -
\alpha^{r\dagger}_{-\mathbf{k},i} \alpha^{r}_{-\mathbf{k},i} \Big)~,
\end{equation}
with  $i = 1,2$. \\ For mixed fields,  one has  $\mathbf{P}_\sigma
(t)=G^{-1}_{\theta}(t) \mathbf{P}_i G_{\theta}(t)$, namely
\be \mathbf{P}_\sigma (t) = \sum_{r=1,2} \int d^3 \mathbf{k} \,
\mathbf{k} \Big( \alpha^{r\dagger}_{\mathbf{k},\sigma}(t)
\alpha^{r}_{\mathbf{k},\sigma}(t) -
\alpha^{r\dagger}_{-\mathbf{k},\sigma}(t)
\alpha^{r}_{-\mathbf{k},\sigma}(t) \Big), \ee
for $\sigma = e,\mu$ with $\mathbf{P}_e(t)+\mathbf{P}_\mu(t)=
\mathbf{P}_1 + \mathbf{P}_2 \equiv \mathbf{P}$ and $[\mathbf{P},
G_{\theta}(t)]=0$. The total momentum is of course conserved, $[{\bf
P}, H]=0$, with $H$ denoting the Hamiltonian. The expectation value on
the flavor vacuum of the momentum operator $\mathbf{P}_\sigma(t)$
vanishes at all times:
\begin{equation}
_{e,\mu}\langle{0 (t)} |\mathbf{P_\sigma (t)}| {0 (t)} \rangle_{e,\mu} = 0 ,~~\sigma=e,\mu~.
\end{equation}
The state $|\alpha^r_{\mathbf{k},e}\rangle \equiv |
\alpha^r_{\mathbf{k},e}(0)\rangle $ is an eigenstate of the momentum
operator $\mathbf{P}_e(0)$ at time $t=0$,
$\mathbf{P}_e(0)|\alpha^r_{\mathbf{k},e}\rangle \equiv \mathbf{k}
|\alpha^r_{\mathbf{k},e}\rangle$.  At time $t\ne0$ the normalized
expectation value for the momentum in such a state is
\bea \mathcal{P}^e_{\mathbf{k},\sigma}(t)&\equiv& \frac {\langle
\alpha^r_{\mathbf{k},e}| \mathbf{P}_\sigma(t)| \alpha^r_{\mathbf{k},e}
\rangle}{\langle \alpha^r_{\mathbf{k},e}| \mathbf{P}_\sigma(0)|f
\alpha^r_{\mathbf{k},e}\rangle} \nn\\ &=& | \{
\alpha^r_{\mathbf{k},e}(t), \alpha^{r\dagger}_{\mathbf{k},e}(t')\}|^2
+ | \{ \alpha^{r\dagger}_{-\mathbf{k},e}(t),
\alpha^{r\dagger}_{\mathbf{k},e}(t')\} |^2~,\nn \eea for
$\sigma=e,\mu$.\\
Note that $\mathcal{P}^e_{\mathbf{k},\sigma}(t)$ behaves actually as a
``charge operator''. Indeed, the operator $
\alpha^{r\dagger}_{\mathbf{k},i} \alpha^{r}_{\mathbf{k},i} -
\alpha^{r\dagger}_{-\mathbf{k},i} \alpha^{r}_{-\mathbf{k},i} $ is the
fermion number operator. Therefore, the explicit calculation of
$\mathcal{P}^e_{\mathbf{k},\sigma}(t)$ provides the flavor charge
oscillation. We obtain
\bea \vspace{-1.3mm} \mathcal{P}^e_{\mathbf{k},e}(t) =
1-\sin^2{2\theta}\nn\\ \times \Big[ |U_{\mathbf{k}}|^{2} \sin^2{\frac
{\omega_{k,2} -\omega_{k,1} }{2} t} +|V_{\mathbf{k}}|^{2} \sin^2{\frac
{\omega_{k,2} +\omega_{k,1} }{2} t} \Big], \nn\\
\mathcal{P}^e_{\mathbf{k},\mu}(t) = \sin^2{2\theta}\nn\\ \times \Big[
|U_{\mathbf{k}}|^{2} \sin^2{\frac {\omega_{k,2} -\omega_{k,1} }{2} t}
+|V_{\mathbf{k}}|^{2} \sin^2{\frac {\omega_{k,2} +\omega_{k,1} }{2} t}
\Big].  \eea Notice that in the absence of the condensate
contribution, i.e. in the $|V_{\mathbf{k}}| \rar 0$ limit
($|U_{\mathbf{k}}| \rar 1$), the usual QM Pontecorvo approximation of
the oscillation formulae is obtained. In the same limit, the
noncommutative structure of the Hopf coproduct algebra (and the
related Bogoliubov transformation) is lost. The quantum field
nonperturbative structure is thus essential for the NCSG construction.


\section{Conclusions}
\label{conclusions}

We have shown that neutrino mixing is {\it naturally} embedded within
the NCSG model. This has been obtained from the doubling of the
algebra $\mathcal{A} = \mathcal{A}_1 \otimes \mathcal{A}_2 $ acting on
the space $\mathcal{H} = \mathcal{H}_1 \otimes \mathcal{H}_2$. In
fact,  by considering the neutrino mixing, we have seen
in Section V that the transformation linking mass annihilation and
creation operators with the flavor ones is a rotation combined
(``nested'') with Bologiubov transformations
(cf. Eqs.~(\ref{opcrea})).  This transformation is the seed of the
mixing annihilation and creation operators leading to the unitarily
inequivalence between the two vacuum states, i.e. mass vacuum and
flavor vacuum. In Section IV we have shown that the Bogoliubov
transformed operators, $a(\Theta)$ and $\tilde{a}(\Theta)$, are linear
combinations of the coproduct operators defined in terms of the
deformation parameter $q(\Theta)$ and its $\Theta$-derivatives,
obtained from the doubled algebra $\mathcal{A} = \mathcal{A}_1 \otimes
\mathcal{A}_2 $. Neutrino mixing is thus intimately related to the
algebra doubling and, as such, it is intrinsically present in the NCSG
of model.

We stress that Bogoliubov transformations act on operators, so our
discussion is framed in the quantum operator formalism. Thus, the
doubling of the algebra in Connes' construction appears to be grounded
in the QFT Hopf deformed algebra, and in turn this has been shown to
involve field mixing. Having to do with fields introduces crucial
features in the formalism. From the one side, it means that we have an
infinite number of degrees of freedom (therefore we have to consider
the continuum or the infinite volume limit). On the other side, as it
emerges from the discussion presented above, the algebra doubling,
through the Bogoliubov transformations, combines the field operator
positive frequency part with the negative frequency one, leading to
the noncommutative features.

It has been shown in Ref.~\cite{Sakellariadou:2011wv} that the gauge
structure of the Standard Model is implicit in the algebra doubling, a
key ingredient of the NCSG construction. In the present paper we have
established the link between the algebra doubling and the field
mixing, concluding that Standard Model derived from the NCSG model,
includes neutrino mixing by construction.


\end{document}